\documentclass{emulateapj}
\usepackage{apjfonts,amsmath}
\include{graphics}
\RequirePackage{epsf,graphicx}

\begin{document}
\title{Galactic constraints on CHAMPs}
\author{F.~J.~S\'anchez-Salcedo
and E.~Mart\'{\i}nez-G\'{o}mez}
\affil{ Instituto de Astronom\'{\i}a, Universidad 
Nacional Aut\'onoma de M\'exico, Ciudad Universitaria,
Apt.~Postal 70 264, C.P. 04510, Mexico City, Mexico}
\email{jsanchez@astroscu.unam.mx, egomez@astroscu.unam.mx}
\begin{abstract}

We improve earlier Galactic bounds that can be placed on the fraction
of dark matter in charged elemental particles (CHAMPs).
These constraints are of interest for CHAMPs whose mass is too large
for them to have seen through their
electromagnetic interaction with ordinary matter, 
and whose gyroradius in the galactic magnetic field is too small 
for halo CHAMPs to reach Earth.  If unneutralized CHAMPs in that
mass range are well mixed in the halo, they 
can at most make up a fraction 
$\lesssim (3-7)\times 10^{-3}$ of the mass of the Galactic halo. 
CHAMPs might still be a solution to the cuspy halo problem if they
decay to neutral dark matter but a fine-tuning is required.
We also discuss the case where CHAMPs do not populate a spherical halo.

\end{abstract}

\keywords{
dark matter -- galaxies: halos -- galaxies:
kinematics and dynamics -- galaxies: magnetic fields }

\section{Introduction}
Massive particles with integer electric charge (CHAMPs), denoted
here by $X$, were considered
as dark matter candidates in the late eighties (De R\'ujula et al.~1990;
Dimopoulos et al.~1990; Gould et al.~1990; Chivukula et al.~1990).
Although stable CHAMPs are predicted in some
extensions of the standard model, astrophysical constraints
plus bounds from underground detectors, 
from balloon experiments and the lack of detection of anomalous
hydrogen in the sea water, basically rule out CHAMPs as dark matter
(see Perl et al.~2001 and Taoso et al.~2008).
In particular, the non-detection of heavy water in the sea excludes
CHAMPs with masses between $10$ and $10^{4}$ TeV (Verlerk et al.~1992).
All the above constraints were derived for the standard flux
of particles at Earth from the Galactic halo. 
If magnetic fields prevent the flux of CHAMPs to
penetrate the Galactic disk, one must reevaluate earlier bounds
(Chuzhoy \& Kolb 2008).

Essentially all $X^{-}$ should have bound to protons,
forming neutraCHAMPs, which 
decouple from the photon-baryon fluid and drive structure formation
prior to recombination. 
NeutraCHAMPs reach Earth unimpeded. Searches for neutraCHAMPs in
cosmic rays rule out particles with masses between $100$ and a few $10^{4}$ 
TeV (Barwick et al.~1990). 
Nevertheless, heavy-water searches, cosmic rays searches, and constraints from
overproduction of $^{6}$Li (Berger et al.~2008) are only relevant
if CHAMPs are singly charged, because for other charges, a CHAMP no longer
behaves like a proton.

If a significant fraction of
the mass of halos is made up by CHAMPs, it may have a strong
impact on the observable Universe (e.g., Chuzhoy \& Kolb 2008).  
It is therefore important to constrain the abundance of CHAMPs
in galactic halos. 
After revising Galactic requirements for CHAMPs to be absent
in the Galactic disk, we give an upper limit on the abundance of
CHAMPs in the Galactic halo.

\section{Shielding the disk with magnetic fields}
\label{sec:shielding}
While neutraCHAMPs have no
difficulties to penetrate the Galactic disk and the solar wind to 
reach Earth, the penetration of unneutralized CHAMPs may be impeded
by the presence of magnetic fields.  
Denote by $\epsilon$ the electric charge of CHAMPs 
in units of $e$, the elementary electron charge.
For $m_{X}>10$ TeV and $\epsilon \leq 1$, CHAMPs in the Galactic
halo behave as a collisionless plasma because the self-collision time 
is $>10^{5}$ Gyr and the mean free path is
$>10^{6}$ kpc. Such a plasma consists of charged particles 
influenced only by gravity and electromagnetic fields. 
We will consider first the interaction of halo CHAMPs with the Galactic 
magnetic field. 

\begin{figure}
\epsscale{1.09}
\plotone{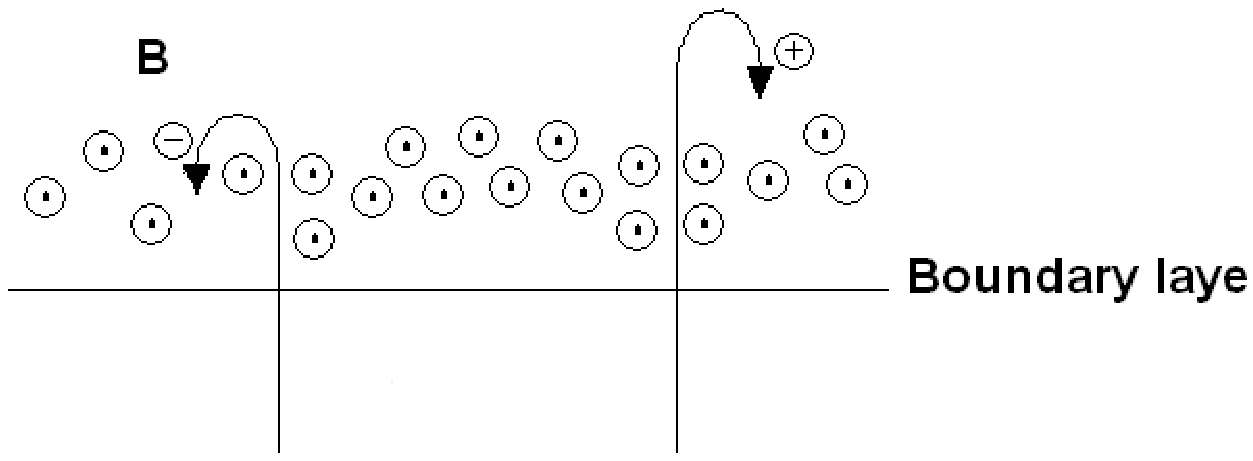}
  \caption{Structure of the magnetic boundary layer formed in the caps
of galactic disks by the partial
penetration of the charged particles before they are deflected back. We also
illustrate the corresponding orbits of particles in the neighborhood of such
boundary for both types of charge, that is, negative and positive.
}
\label{fig:boundary}
\end{figure}

It is well-known that when charged particles 
interact with a magnetized body, a boundary
layer that divides two regions with different conditions is created (Parks
1991). Thus, charged particles will penetrate this boundary by some
distance before they are turned around by the $\vec{v}\times \vec{B}$ 
force (Figure \ref{fig:boundary}). 
The boundary layer is formed because of the partial penetration
of the charged particles before they are deflected back.
The orbits described by negative and positive charged particles in
the neighborhood of the magnetic boundary are drawn in Fig.~\ref{fig:boundary}.

Since the Galactic magnetic field is not perfectly plane-parallel and has a
nonzero turbulent component, i.e.~$\vec{B}=\vec{B}_{0}+\vec{b}$, where $\vec{%
B}_{0}$ is the regular (homogeneous) magnetic field and $\vec{b}$ denotes
the turbulent field, the
particle motion is determined not only by the average magnetic field but
also by scattering at field fluctuations, a stochastic process which
requires the solution of transport equations with particle ensembles.
Depending on the magnitude of these fluctuations, we distinguish between
weak and strong turbulence which leads to different physical phenomena
(Kallenrode 1998).
Particle propagation in turbulent fields can be understood as a diffusive
process, reason why we consider the spatial diffusion of \textit{%
collisionless} halo CHAMPs into the galactic disk.
The diffusion timescale $\tau_{\rm diff}$ accross the galaxy 
disk thickness $H$ for a halo CHAMP, is bracketted in the range:
\begin{equation}
\frac{H^{2}}{2D_{\parallel}}<\tau_{\rm diff}\lesssim \frac{H^{2}}{2D_{\perp}},
\end{equation}  
where $D_{\parallel}$ and $D_{\perp}$ are the diffusion
coefficients parallel and transverse to the mean component of the magnetic
field, which is observed to be parallel to the disk. 
Within the disk, the magnetic field can
be considered static because Alfv\`en waves propagate with velocities of the
order of the Alfv\`en speed $v_{A}\sim 6$ km s$^{-1}$, which is smaller 
than the typical velocities of CHAMPs $\sim \sqrt{3}v_{X}$, where
$v_{X}\simeq 150$ km s$^{-1}$ is the one-dimensional velocity dispersion 
for halo particles.
The diffusion coefficients depend on the turbulence level $\eta\equiv
(1+\left<B_{0}^{2}\right>/\left<b^{2}\right>)^{-1}$, 
and on the rigidity $\chi\equiv 2\pi r_{L}/\lambda_{\rm max}$,
with $r_{L}$ the Larmor radius defined with respect to the total
magnetic field and $\lambda_{\rm max}$ the maximum
scale of the turbulence $\sim H/2$ (Giacalone \& Jokipii 1999;
Casse et al.~2002).
Observations of the Galactic polarized synchrotron background yield 
$1<\left<b^{2}\right>/\left<B_{0}^{2}\right><9$ 
(Fletcher \& Shukurov 2001, and references therein),  
implying that $0.5<\eta<0.9$.
Since $\tau_{\rm diff}$ scales as the inverse of the
diffusion coefficients and those are essentially a
monotonic function of $\eta$, we use $\eta\simeq 0.5$ in our
estimate of $D_{\perp}$ in order to give an upper limit on the
diffusion timescale.
Taking advantage of the numerical result by Casse et al.~(2002)
that $D_{\perp}/(r_{L}v)\sim 0.3$ for Kolmogorov turbulence with
$\eta =0.5$ and $\chi$ between $0.05$ and $0.4$,
we find that
\begin{eqnarray}
\tau_{\rm diff}\lesssim & & \frac{5H^{2}}{3r_{L}v_{X}}
=9\epsilon \,{\rm Gyr} \left(\frac{H}{300{\rm pc}}\right)^{2}
\left(\frac{m_{X}}{10^{6}{\rm TeV}}\right)^{-1}
\nonumber\\
&&
\times \left(\frac{v_{X}}{150{\rm km s}^{-1}}\right)^{-2}
\left(\frac{B}{5\,\mu{\rm G}}\right).
\end{eqnarray}
Therefore, the present configuration and strength of the Galactic
magnetic field can
prevent diffusion of (unaccelerated) CHAMPs across the Galactic disk 
in the life time of the disk
for mass particles $m_{X}< 10^{6} \epsilon$ TeV.
The corresponding gyroradius for a mass of $10^{6}\epsilon$ TeV
moving at $150$ km s$^{-1}$ in a field of $5\mu$G is $0.1$ pc.

It is likely that CHAMPs are accelerated to much higher velocities
by supernova shocks. The inclusion of this effect would give a more
constraining upper limit on $m_{X}$. 

\section{Energy loss of CHAMPs in the disk}
\begin{figure}
\epsscale{1.09}
\plotone{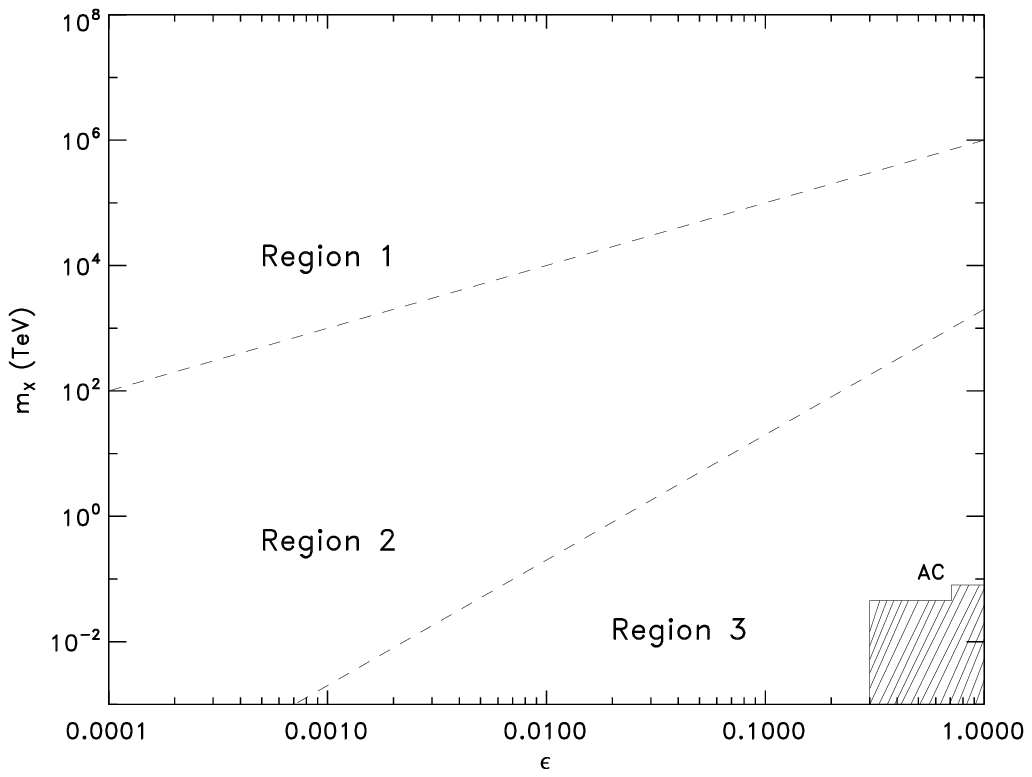}
  \caption{Regions of mass-charge space of astrophysical relevance.
Halo milli-charged particles in region ``1''  
have no difficulty in penetrating the disk, whereas CHAMPs 
in region ``3'' would
interact strongly with the baryonic matter in the disk resulting in strong
observable consequences, unless its abundance is very tiny.
For reference, the excluded region
from accelerator experiments (AC) after Davidson et al.~(2000), is 
also shown.}
\label{fig:regions}
\end{figure}

In the preceding section we have seen that halo CHAMPs
with masses $m_{X}< 10^{6} \epsilon$ may have difficulty
penetrating the magnetized Galactic disk, whereas those inside it
stay confined to the disk.  Trapped particles in the disk
gain energy through electrostatic fields, Fermi acceleration in shock waves, 
and its descendants (e.g., Blandford 1994),
and loss kinetic energy due to Coulomb scatterings
with electrons and protons of the diffuse interstellar gas. 
The dissipation timescale is $\tau_{\rm dis}=E/|\dot{E}|$, with
$E=mv_{X}^{2}/2$ and 
\begin{equation}
|\dot{E}|=4\pi n_{e}\frac{\epsilon^{2}e^{4}}{m_{e}v_{X}}\ln\Lambda,
\end{equation}
where $n_{e}$ is the electron density ($\approx 0.025$ cm$^{-3}$ in the
solar vicinity) and the Coulomb logarithm has a value of
about $20$.  CHAMPs may avoid strong cooling 
if the dissipation timescale $\tau_{\rm dis}$ for
CHAMPs trapped on the disk be greater than the shock acceleration
timescale $\tau_{\rm acc}$. Since $\tau_{\rm acc} \gtrsim 0.01$ Gyr,
the condition $2\tau_{\rm cool}>\tau_{\rm acc}=0.01$ Gyr implies 
\begin{equation}
m_{X}> 2\times 10^{3}\epsilon^{2} \, {\rm TeV} 
\left(\frac{v_{X}}{150{\rm km s^{-1}}}\right)^{-3}.
\label{eq:cooling}
\end{equation}
This constraint is valid for any value of $\epsilon$ provided
that $m_{X}$ is larger than the electron mass $m_{e}\sim 0.5$ MeV.
We will not consider the regime $m_{X}<m_{e}$ because they are
excluded for $10^{-15}\lesssim \epsilon <1$ 
(Davidson et al.~2000).
In Fig.~\ref{fig:regions} we plot the region of mass-charge space 
for CHAMPs, combining the constraint derived in \S \ref{sec:shielding}
and Eq.~(\ref{eq:cooling}).  CHAMPs in region ``2''
can be suspended in the halo without penetrating
the disk and may be very evasive for direct terrestrial detection because 
the flux of CHAMPs reaching Earth may be highly suppressed.
If CHAMPs are thermal relics, they must also
satisfy the  unitary bound $m_{X}\lesssim 120$ TeV
(Griest \& Kamionkowski 1990). Combining this constraint with 
Eq.~(\ref{eq:cooling}), we find $\epsilon<0.2$ for thermal relics.

\section{The global magnetic support}
For simplicity, let us assume for a moment that the Galactic magnetic field
is horizontal.  As depicted in Fig.~\ref{fig:boundary},
CHAMPs with $m_{X}< 1\times 10^{6}\epsilon$ TeV 
execute approximately half a 
gyro-orbit before finding themselves back in the unmagnetized region 
and with velocities directed away from the magnetized region. 
This leads to the plasma being excluded from the magnetized disk.
In the boundary layer, a current
layer develops as a result of a thermal, unmagnetized plasma interacting 
with a magnetized region. 
It is a classical result that the (kinetic) motions of individual 
particles in collisionless 
plasmas can be reconciled with the role inferred for the pressure in MHD
(e.g., King \& Newmann 1967; Cravens 1997). 
For example, in a boundary layer like the terrestrial magnetopause,
the thermal (or kinetic) pressure of the solar wind is balanced
by the pressure of the terrestrial magnetic field and this separates
the interplanetary magnetic field from the magnetospheric cavity.

Suppose that CHAMPs cannot penetrate
down to $z=Z_{\rm min}$ because of the Galactic magnetic barrier 
(see \S \ref{sec:shielding}). 
Since CHAMPs are essentially collisionless, the magnetic field
is the only agent that may support the weight of the CHAMPs in the halo.
Integrating the equation of vertical
equilibrium from $z=Z_{\rm min}$ to $z=\infty$, assuming zero pressure
at $z=\infty$, and ignoring the weight of coronal gas, we find
\begin{equation}
P_{B}(Z_{\rm min})\gtrsim \int_{Z_{\rm min}}^{\infty} \rho_{ch} K_{z} dz,
\label{eq:graveq}
\end{equation}
where $P_{B}$ is the magnetic pressure, $\rho_{ch}(z)$ the mass density
of CHAMPs and $K_{z}$ the vertical positive gravitational
acceleration.  For a spherical dark halo,
the contribution of CHAMPs to the weight term at the solar vicinity is
\begin{eqnarray}
\int_{Z_{\rm min}}^{\infty} \rho_{ch} K_{z} dz= && 
f\rho_{X}v_{X}^{2}= 
1.7f\times 10^{-10}{\rm dyn}\,\, {\rm cm}^{-2}
\nonumber \\ &&
\times \left(\frac{\rho_{X}}
{0.01{\rm M}_{\odot}{\rm pc}^{-3}}\right)
\left(\frac{v_{X}}{150\;{\rm km} \;{\rm s}^{-1}}\right)^{2},
\end{eqnarray}
where $f$ is defined as the mass fraction of dark matter in CHAMPs. 
The weight of the CHAMPs produces a 
confinement effect. Interestingly, the observed synchrotron emission
above the plane in the solar neighbourhood implies that the scale height
of the magnetic field is greater than what would be inferred from 
the weight distribution of the interstellar matter (e.g., Cox 2005).
By requiring that the confinement of the magnetic pressure is 
entirely due to the weight of the CHAMPs,
an upper value on the abundance of CHAMPs can be derived.

The observed synchrotron
emission above the plane in the solar neighbourhood indicates that
the total magnetic field strength is
$2-5\, \mu$G at a height of $z=1$ kpc 
(Ferri\`ere 2001; Cox 2005; Gaensler et al.~2008). 
If we identify $Z_{\rm min}$ as the HWHM of the magnetoionic disk
$\sim 1$ kpc (e.g., Kalberla 2003) and
by equating the magnetic pressure at $z=Z_{\rm min}\approx 1$ kpc
with the weight term, 
we obtain the desired constraint on $f$, once
adopting the highest magnetic value of $5\mu$G allowed by observations:
\begin{equation}
f\leq  7\times 10^{-3} 
\left(\frac{\rho_{X,\odot}}
{0.01{\rm M}_{\odot}{\rm pc}^{-3}}\right)^{-1}
\left(\frac{v_{X}}{150\;{\rm km}\;{\rm s}^{-1}}\right)^{-2}.
\end{equation}
This constraint is independent of charge $\epsilon$.
This estimate is very robust to 
the precise value adopted for $Z_{\rm min}$ because the magnetic field
decays very slowly with $z$.

In our derivation, we have assumed that the halo is spherical. 
Consider now an oblate isothermal dark halo with axis ratio $q$:
\begin{equation}
\rho_{X}(R,z)=\frac{v_{c}^{2}}{4\pi G\alpha q}
\left(R^{2}+\frac{z^{2}}{q^{2}}\right)^{-1},
\end{equation}
where $v_c$ is the asymptotic circular velocity at the equatorial plane
and $\alpha=\gamma^{-1}\arcsin \gamma$, with $\gamma=\sqrt{1-q^{2}}$.
In this model, the velocity dispersion is given, within
less than $10\%$, by 
$v_{X}\simeq 1.16 \sqrt{q}(v_{c}/\sqrt{2})$, for flattening $0.05<q<0.5$ 
(e.g.~Gerhard \& Silk 1996).
Even though the velocity dispersion for $q<1$ is smaller than
in the spherical case, the weight term 
changes only by $\sim 10\%$ as compared to the spherical case,
even for rather flattened halos ($q\approx 0.5$).

So far, it was assumed that CHAMPs and neutraCHAMPs are well mixed.
Now, we relax that assumption and search for the distributions of neutraCHAMPs
and CHAMPs that allow to have the largest mass in CHAMPs. 
This occurs when CHAMPs have an almost zero vertical
dispersion and settle on to layers of negligible thickness at 
$|z|=Z_{\rm min}$, but neutraCHAMPs populate a spherical halo. 
The vertical density profile of CHAMPs is then
$\rho_{ch}(z)=\Sigma_{ch}\delta (|z|-Z_{\rm min})/2$, 
where $\Sigma_{ch}$
is the surface density of CHAMPs at $R=R_{\odot}$. From Eq.~(\ref{eq:graveq}),
we obtain that $P_{B}\gtrsim\Sigma_{ch} K_{z}/2$ at $z=Z_{\rm min}$. Taking 
$K_{z}\approx 6\times 10^{-9}$ cm s$^{-2}$ at $z=1$ kpc (e.g., Holmberg
\& Flynn 2004), we find $\Sigma_{ch}<6$ M$_{\odot}$ pc$^{-2}$ or, 
equivalently,
$f\approx \Sigma_{ch}/(2\rho_{X,\odot}R_{\odot})\lesssim 3.5\times 10^{-2}$. 
However, it is difficult to justify this additional degree of
freedom in the model until a non-gravitational mechanism for so 
efficiently dissipation of CHAMP's energy is satisfactorily established.
In addition, it is likely that Parker instabilities will destroy these
cold layers of CHAMPs if they are sustained against gravity by 
magnetic fields. 

Consider now a portion 
of the disk at larger galactocentric distances, say $R=2R_{\odot}$.
Following the same procedure than in the solar neighbourhood,
we need to estimate the total magnetic pressure at 
$(2R_{\odot},Z_{\rm min})$, 
which should be responsible to give support
to the halo CHAMPs.  The large-scale magnetic field
may have a scaleheight $5$--$10$ times the scaleheight of the neutral
gas disk, so that we may assume that $B_{0}(Z_{\rm min})\simeq
B_{0}(z=0)$. 
The random magnetic field is expected to
be roughly in equipartition with the kinetic energy in the turbulence.
Therefore,
its vertical scaleheight should be similar to that of the gas. If
magnetic fields are still a barrier for halo CHAMPs,
then we may assume that $Z_{\rm min}>H$
and, consequently, the magnetic pressure by the random component
at $Z_{\rm min}$ is less than $10\%$ the pressure by the random field
at $z=0$. Collecting both contributions, we derive an
upper limit for the total magnetic pressure
at $Z_{\rm min}$:
\begin{equation}
P_{B}<\frac{B_{0}^{2}+0.1b^{2}}{8\pi}
=\frac{B_{0}^{2}}{8\pi}(1+0.1\alpha),
\end{equation}
where $\alpha\equiv b^{2}/B_{0}^{2}$, with $b^{2}$ and 
$B_{0}^{2}$ evaluated at $z=0$.
The ordered magnetic field is difficult to measure in the outer
Galaxy, but there is evidence that it decays with radius $R$
as a power-law between
$R^{-1}$ and $R^{-2}$, probably as $\exp(-R/R_B)$ with $R_{B}=8.5$ kpc
(Heiles 1996; Han et al.~2006).
The uniform magnetic field in the  solar neighbourhood is
$2$--$4\mu$G, depending on the authors (Beck 2002; Han et al.~2006).
If we generously take a value in the solar circle of $4\mu$G
we infer a strength $B_{0}\sim 1.5\mu$G at $2R_{\odot}$.
Assuming a spherical dark halo with a mass density at $2R_{\odot}$ of 
$\sim \rho_{X,\odot}/4$, then 
$\int \rho_{ch} K_{z} dz 
=f\rho_{X,\odot}v_{X}^{2}/4$. 
At $2R_{\odot}$, our assumption that the halo
is spherical is a very good approximation (e.g., Belokurov et al.~2006;
Fellhauer et al.~2006).
By imposing pressure balance at $z=Z_{\rm min}$ (Eq.~\ref{eq:graveq}),
the following constraint for $f$ is inferred
\begin{equation}
f\leq 2\times 10^{-3} \left(1+0.1\alpha\right)
\left(\frac{\rho_{X,\odot}/4}
{0.0025{\rm M}_{\odot}{\rm pc}^{-3}}\right)^{-1}
\left(\frac{v_{X}}{150\;{\rm km}\;{\rm s}^{-1}}\right)^{-2}.
\label{eq:twicesolar}
\end{equation}
Other observational estimates assure our
generously-taken magnetic intense.  In fact,
data from rotation
measurements of pulsars suggest uniform magnetic fields of
$\sim 0.7\,\mu$G at $R=2 R_{\odot}$ (Rand \& Lyne 1994), which coincides with
the extrapolation of the fit of radial variation of the regular field 
by Han et al.~(2006).

Beyond $2R_{\odot}$ it is uncertain if supernovae shocks are able
to clean the disk from CHAMPs. It might be also possible
that beyond the optical radius, 
the magnetic field is too weak to prevent CHAMPs from crossing the disk,
but any more complicated analysis is useless in the face of such
ignorance.  We conclude that charged particles 
can be suspended in the halo, so that they 
would be impossible to detect as they never reach the Earth. 
However, the mass fraction of bare CHAMPs in the halo
must be rather small $f\lesssim (2-7)\times 10^{-3}$. 
In a $X^{+}$-$X^{-}$ symmetric Universe,
neutraCHAMPs may compose a fraction $< (2-7)\times 10^{-2}$,
because the ratio of their relative abundances may be as much as 10:1
(Dimopoulos et al.~1990). 
Therefore, CHAMP models require three ingredients:
neutraCHAMPS, CHAMPS and neutral dark matter. One possibility is that
$f\approx 0.5$ \footnote{We adopt $f=0.5$ because
positive CHAMPs and neutraCHAMPs
are expected to be approximately in equal numbers.} at early times and 
CHAMPs decay to neutral dark matter with a lifetime 
$<2.75$ Gyr to reach $f\sim 3\times 10^{-3}$ at present.

\section{Discussion}

\subsection{The origin of cores in galaxies}
Our upper limit on $f$ rules out
CHAMPs as an explanation for the formation of constant density cores in
dark matter halos. The reduction of the central density 
after they have driven the formation of galactic halos would be
insignificant. Even if all the CHAMPs were depleted from
the central parts of the galaxies, the rotation velocity in a certain
galaxy would suffer a negligible change of $(0.1-0.35)\%$ for
$f\sim (2-7)\times 10^{-3}$. If CHAMPs decay to neutral particles,
$f$ was larger in the past. In such scenario, 
the formation of cores may result from
the evacuation of CHAMPs in the central regions of galaxies by supernova 
shocks, provided that the decay times $\gtrsim 1.5$ Gyr.
Combined with the constraint that the decay times $<2.75$ Gyr, discussed
in our previous section, a fine-tuning of the decay time is
required.

\subsection{Ram pressure stripping and collisions of galaxy clusters} 
Magnetic fields couple CHAMPs with themselves and with ordinary matter.
This coupling might 
cause ram pressure stripping of both baryonic and dark matter of 
subhalos and satellite systems.
Consider, for instance, the collision of two galaxy clusters.
Estimates for the magnetic field strength in clusters range from roughly
$1-10\,\mu$G at the center and $0.1-1\,\mu$G at a radius of $1$ Mpc,
values that correspond to a plasma beta for the CHAMPs, 
$\beta\equiv 8\pi P_{th}/B^{2}\approx 2f\times 10^{3-4}$, a hot plasma.
Even in this dynamically weak magnetic field,
the mean gyroradius for a CHAMP with $m_{x}=10^{6}\epsilon$ TeV,
is $\lesssim 5$ pc at the center and 
$\lesssim 50$ pc at $1$ Mpc.
The governing equations of collisionless hot plasmas were developed
by Chew et al.~(1956), whose theory is known as the Chew-Goldberger-Low
approximation. This approximation, which leads to MHD equations with
anistropic pressure, is satisfactory when the Larmor frequency is
large compared to other characteristic frequencies of the problem and
the mean particle gyroradius is short compared to the distance
over which all the macroscopic quantities change appreciably
(e.g., Spitzer 1962; Schmidt 1966).
Therefore, charged massive particles in the halo of galaxy clusters  
can be described in the
fluid-like anisotropic MHD approximation; in the merger process,
they would behave as a clump of fluid, experiencing ram pressure 
stripping and drag deceleration similar to the gas component.
Since CHAMPs should be attached to the gas component,
the observed offset between the centroid of dark matter 
and the collisional gas of the subcluster
in the Bullet Cluster implies $f\ll 1$
(e.g., Natarajan et al.~2002; Markevitch et al.~2004). 
Although the current lensing data accuracy is not sufficient to derive
the mass distribution of the subcluster in the Bullet Cluster, the derived mass
estimates of the subcluster leave little room for 
dark matter in the gas bullet.

Galactic halo CHAMPs 
may also exert ram pressure on the gas component of the LMC and its stream
due to their continuous scattering by
the intrinsic magnetic field of the LMC and the Magellanic stream. 
For a Milky Way-type halo of $\sim 10^{12}$ M$_{\odot}$, 
a fraction $f$ of $3\times 10^{-3}$
implies that the mass in CHAMPs could be up to $\sim 3\times 10^{9}$
M$_{\odot}$ and the density at $50$ kpc of $0.9\times 10^{-6}$ M$_{\odot}$
pc$^{-3}$.  Since these values are smaller
than those required to explain the mass and extension of the Magellanic
Stream and the size and morphology of the gaseous disk of LMC
(Mastropietro et al.~2005),
we cannot reduce any further our upper limit on $f$
with the current observations of the LMC disk and the Magellanic Stream.

\section{Conclusions}

Whilst the common wisdom holds that dark matter is neutral and collisionless,
it is important to explore the possibility of it having nonzero,
not necessarily integer, charge.
We have considered the pressure support of CHAMPs 
in our Galaxy to derive a simple, upper limit
on the fraction of CHAMPs and milliCHAMPs, $f\lesssim (2-7)\times 10^{-3}$.
If $f$ was roughly constant over time,
this constraint rules out CHAMPs as the origin of the cores in LSB
and dwarf galaxies.
In the range of astrophysical interest,  
CHAMPs behave like strongly interacting (fluid-like) dark matter (SIDM).
Thus, they face many of the problems attributed to SIDM.
As some examples, we have discussed the survival of the Magellanic Stream and 
the mass distribution of the
Bullet Cluster.
Our constraint that the mass in CHAMPs in the Galaxy is not larger than the
mass of coronal gas in the halo seems to apply also to galaxy clusters.

\acknowledgments
We are indebted to Leonid Chuzhoy for valuable comments which
led to significant improvements in the manuscript.
We are grateful to Julio Martinell for many important discussions.
F.J.S.S.~acknowledges financial support from PAPIIT project
IN114107 and CONACyT 2006-60526.
E.M.G.~thanks support from DGAPA-UNAM postdoctoral fellowship.

\end{document}